\documentclass[aps,prl,twocolumn,showpacs,preprintnumbers]{revtex4-1}

\usepackage{psfrag,graphicx}
\usepackage{dcolumn}
\usepackage{amsmath,amssymb}
\usepackage{bm}
\usepackage{amsfonts,amssymb,amsmath}        % for math symbols.
\usepackage{epstopdf}
\usepackage{mathtools}
\DeclarePairedDelimiter{\ceil}{\lceil}{\rceil}

\newcommand{\be}{\begin{equation}}
\newcommand{\ee}{\end{equation}}
\newcommand{\bq}{\begin{eqnarray}}
\newcommand{\eq}{\end{eqnarray}}

\newcommand{\ket}[1]{\left | \, #1 \right\rangle}

\begin{document}
\title{A repetition code of 15 qubits}
\author{James R. Wootton and Daniel Loss}
\affiliation{Department of Physics, University of Basel, Klingelbergstrasse 82, CH-4056 Basel, Switzerland}

\begin{abstract}

The repetition code is an important primitive for the techniques of quantum error correction. Here we implement repetition codes of at most $15$ qubits on the $16$ qubit \emph{ibmqx3} device. Each experiment is run for a single round of syndrome measurements, achieved using the standard quantum technique of using ancilla qubits and controlled operations. The size of the final syndrome is small enough to allow for lookup table decoding using experimentally obtained data. The results show strong evidence that the logical error rate decays exponentially with code distance, as is expected and required for the development of fault-tolerant quantum computers. The results also give insight into the nature of noise in the device.

\end{abstract}


\maketitle

\section{Introduction}

The development of quantum computing is entering an exciting new era. Prototype quantum processors based on various physical architectures are beginning to appear, such as those superconducting qubits \cite{kelly:14,takita:16}, trapped ions \cite{linke:16} and spin qubits \cite{watson:17}. Such devices are too small and noisy to fully realize the promise of quantum computation. Many more qubits and the ability to achieve fault-tolerance are required before the age of quantum computers truly dawns.  Nevertheless, current and near future devices can and have been used to generate many significant proof-of-principle results.

An important challenge of this new era is to benchmark and compare quantum processors. A quantitative starting point for this is the 'Quantum Volume'\cite{quantumvolume:17}, designed to capture not only the number of qubits in a device, but also some idea of the circuit depth that can be achieved before the effects of noise dominate.

More detail on the capabilities of a device can be obtained by running simple programs. An obvious choice would be to implement small instances of algorithms intended for large fault-tolerant devices, such as Shor's~\cite{shor:97} or Grover's~\cite{grover:96} algorithms. However, a better insight would arguably come from algorithms that have been specifically designed to work on small and noisy devices \cite{awesomeness}.

Fortunately, there are already a class of protocols designed specifically for noisy systems: those of quantum error correction \cite{lidar:13}. Many experiments have already been done based on tasks in this area, such as error detection~\cite{linke:16,vuillot:17,takita:17}, correction~\cite{kelly:14,ofek:16} and proof-of-principle tests of the required techniques~\cite{blatt:14,takita:16,wootton:majorana}.

Several of these experiments have involved large Hilbert spaces. However, most of these have either not allowed both detection and correction \cite{blatt:14}, or not used the standard paradigm of encoding in a many qubit system \cite{ofek:16}. Apart from these, the largest number of qubits used so far for the detection and correction of errors was a repetition code of 9 qubits~\cite{kelly:14}. This code is capable of both detecting and correcting errors on a logical \emph{bit} value stored in an array of qubits. Note that, unlike most quantum error correcting codes, this does not allow a logical \emph{qubit} to be stored. However, the connectivity required to implement codes which store qubits is beyond most current devices~\cite{gambetta:17,bermudez:17}.

Specifically, if the qubits of a device are represented as the vertices of a graph, and edges are placed between all pairs for which entangling gates can be directly performed, full quantum error correcting codes require this graph to be at least a two dimensional planar lattice. Repetition codes, however have much more amenable needs: All they require is a line. This code therefore represents the forefront of current implementations of quantum fault-tolerance, and is an important means to benchmark current devices.

\begin{figure}[t]
\begin{center}
{\includegraphics[width=\columnwidth]{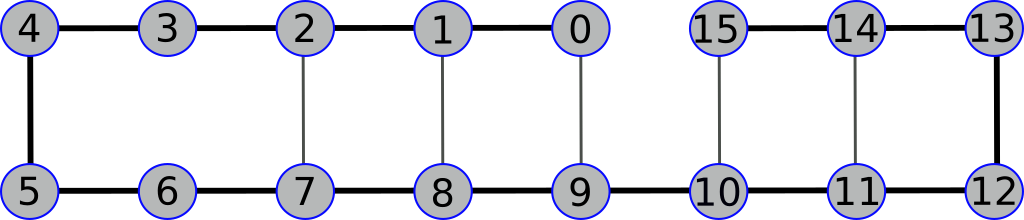}}
\caption{\label{device} The layout of the \emph{ibmqx3} device, with the numbering of qubits used in this study. Lines connect pairs of qubits for which a CNOT can be performed. Thick black lines are show for the CNOTs used in our implementation of the repetition code.
}
\end{center}
\end{figure}

At the time producing this study, the largest quantum processor was the \emph{ibmqx3} of IBM \cite{ibmqx3}. This is a 16 qubit device whose connectivity is described by the $2 \times 8$ square lattice shown in Fig.~\ref{device}. Since repetition code can only be defined on an odd number of qubits, this device can be used to implement a repetition code of up to 15 qubits. It is such an implementation that we consider in this work. By doing so, we can look for evidence of one of the key assumptions and requirements of quantum error correction, namely that the logical error rate decays exponential with code distance.

This exponential decay is typically dependent on another assumption: that noise acts with finite probability on only a finite number of qubits. This could be violated by sufficiently large correlations between code qubits \cite{fowler:correlated,hutter:correlated}, or by the operations used to implement the code inadvertently allowing noise to spread \cite{helsen:18}. In such cases, the noise threshold required for successful error correction would become zero, significantly limiting the degree to which errors can be suppressed. Such effects could be effectively ruled out by demonstrating that the exponential decay of logical error rate persists to arbitrarily large code distances. In this study we will consider this to a limited extent by considering a range of possible code sizes, but it is beyond the scope of this work to fully rule out such effects.

\section{The Repetition Code}

The repetition code stores classical information through repetition. A code with distance $d$ stores the bit value $b \in \{0,1\}$ by simply repeating the value $b$ across $d$ bits.

For example, this encoding step correponds to the following for $d=5$,
\bq \nonumber
0 \rightarrow 00000, \,\,\, 1 \rightarrow 11111.
\eq

To implement this code using a quantum device, we simply replace the bits with qubits. The value $0$ is then stored with the state $\ket{0}^{\otimes d}$, and $1$ is stored with $\ket{1}^{\otimes d}$.

For the example of $d=5$, this is then,
\bq \nonumber
0 \rightarrow \ket{00000}, \,\,\, 1 \rightarrow \ket{11111}.
\eq

If no errors occurred, the result at output would be to simply regain the initial $\ket{0}^{\otimes d}$ or $\ket{1}^{\otimes d}$. Decoding the stored bit value in this case would trivially be an inversion of the encoding, such as,

\bq \nonumber
\ket{00000} \rightarrow 0, \,\,\, \ket{11111} \rightarrow 1.
\eq

When errors do occur, the result will most likely be a bit string with a mixture of $0$s and $1$s. Decoding is then  typically done by majority voting: If the error rate is low, the majority of qubits can still be expected to be correct. Deducing that the majority value is the one that was encoded will therefore allow the encoded information to be retrieved in most cases. The probability that the decoding is incorrect is known as the logical error probability.

For example, suppose errors are bit flips which occur with probability $p<0.5$. If the result $\ket{01000}$ is obtained for $d=5$, it could have resulted from two possible processes. One is that a logical $0$ was encoded, followed by an error on the second qubit. The other is than a $1$ was encoded, and errors occurred on all but the second qubit. The probability of the former is much higher than that of the latter, due to the much smaller number of errors. So the decoding would be $\ket{01000} \rightarrow 0$.

For the error rate to be low enough for good decoding to be possible, the stored information must be retrieved quickly after being first encoded. However, error correction typically aims to store information over long time scales. When the repetition code is implemented classically, this can be achieved by periodically measuring the bits of the code to keep track of errors as they occur.

This tactic is not compatible with the needs of quantum error correction, in which our techniques should not be allowed to collapse a superposition of a stored $0$ and a stored $1$, which would be encoded in the code qubits as $\alpha\ket{0}^{\otimes d}+\beta\ket{1}^{\otimes d}$. We therefore need a corresponding method that detects errors, but does not need to readout the stored bit at the same time. This process corresponds making to so-called `stabilizer measurements'\cite{gottesman:96}, which lie at the heart of most prominent quantum error correcting codes.

\begin{figure}[t]
\begin{center}
{\includegraphics[width=7cm]{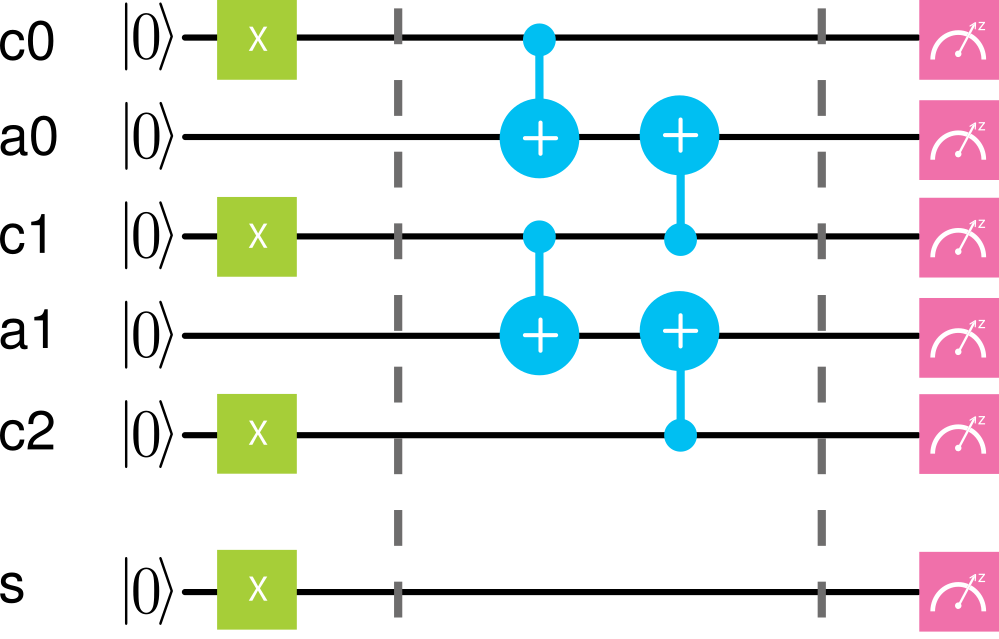}}
\caption{\label{circuit} The basic circuit for a $d=3$ repetition code. 
The case shown is that for which the encoded logical bit value is $1$. This value is encoded in the three code qubits ($c0$, $c1$ and $c2$) by rotating the initial $\ket{0}$ value to $\ket{1}$ using an $X$ gate. The case for an encoded logical $0$ is the same, but without the $X$ gates. The qubit $s$ is not part of the code, but is used to compare the code against the case of encoding the same bit value in a single qubit. The qubits $0$ to $5$, according to the numbering of Fig. \ref{device}, are used to implement this code.
}
\end{center}
\end{figure}

Stabilizer measurements for the repetition code are achieved using the circuit of Fig. \ref{circuit}. In addition to the $d$ code qubits, $d-1$ ancilla qubits are used. The code and ancilla qubits are arranged alternately on a line. Each pair of neighbouring code qubits therefore always has an ancilla located between them. CNOT gates are then performed, with a code qubit as control and an ancilla qubit as target in each case. For each ancilla, a CNOT is applied with both neighbours. The end effect of this on the ancilla does not depend on the encoded bit value. It depends only on whether the computational basis states of the neighbouring code qubits agree (as they should due to the repetition) or not (a signature of error). By measuring the ancillas we can then extract information about errors, without collapsing any encoded superposition.

This portion of the circuit (CNOTs and ancilla measurements) can be repeated indefinitely to keep track of errors as they arise. When it is time to read out the bit value, the code qubits should also be measured. The entire output can then be used to deduce the original intended value. As long as the probability of error between ancilla measurement rounds is sufficiently low, the probability of a logical error will decay exponentially as $d$ is increased.

The decoding in this case can no longer be done simply using majority voting. Instead a decoding method such as the Blossom  algorithm for minimum weight perfect matching can be used \cite{dennis}, as can other methods designed for the case of the surface code with perfect syndrome measurements~\cite{hutter:14,bravyi:14,duclos:14}.

The use of such algorithms is not required if the output is sufficiently small (due to small $d$ and few ancilla measurement rounds). In such cases, a look-up table decoder can be calculated from experimental data.

Specifically, the look-up tables are conditional probability distributions $\pi(R|E)$, where $R$ is a bit string of respresenting the final output of a code, and $E \in \{0,1\}$ is the encoded bit value. These distributions can be determined experimentally for codes of limited size using the results of many runs. Once known, they can be used to decode an arbitrary output $R$. This is done simply by finding the value of $E$ for which $\pi(R|E)$ is highest. This then gives the most likely value for the encoded bit, assuming that the priors for the two values are equal, and so is taken to be the decoded bit value.

The probabilities for logical errors can be determined from the same look-up tables used for the decoding. For a given encoded bit value $E$, the probability $P(R|E)$ that the result $R$ occurs and causes a logical error is clearly

\bq \label{conditional}
P( R|E ) = \begin{cases}
\pi(R|E), & \text{if $\pi(R|\neg E)>\pi(R|E)$}.\\
\pi(R|E)/2, & \text{if $\pi(R|\neg E)=\pi(R|E)$}.\\
0, & \text{otherwise}.
\end{cases}
\eq

The second case here reflects the possibility that the decoding is ambiguous, and so the decoded value is chosen randomly. 

The total probability for a logical error for a given encoded bit value $E$ can then be obtained by summing over all possible outcomes

\bq
P(E) = \sum_R P(R|E).
\eq

In our experiment we consider codes with a maximum of $d=8$ (and so $15$ qubits in total). Also, due to restrictions of the API for the IBM Quantum Experience~\cite{ibmapi}, it is not possible to measure the ancilla qubits repeatedly. We therefore do only a single round of CNOTs and ancilla measurements, the latter of which are done simultaneously with the final readout of code qubits. The circuit applied is therefore exactly that shown in Fig. \ref{circuit} for $d=3$, or generalizations thereof for higher $d$. Due to this limited size, we perform the lookup table decoding described above.

Note that the limited sample size will lead to these results being approximate. Specifically, the method stated above will strictly function as a lower bound due to the possibility of overfitting. An alternative method producing an upper bound is investigated in an alternative version of the source code for this study, which can be found at \cite{repetitionsource}. This is found to give the same qualitative features as the results presented in this work, but is quantitatively different.

\section{Conditions For Success}

In order to determine how well the device implements the repletion code, concrete conditions for success must be defined.

The most straightforward is to compare a bit stored in a code with one stored in only a single qubit. The promise of quantum error correction is that the non-local storage of information across many qubits is more reliable than the single qubits themselves. We must therefore confirm that this the case. To do this, each run of the repetition code with a given encoded bit value is accompanied by a single qubit in which the same value is encoded. This additional qubit, which we will refer to as qubit $s$, is shown at the bottom of Fig. \ref{circuit}.

Another obvious test is to ensure that the probability of a logical error does indeed decrease with $d$, as expected. This should certainly be the trend for large increments $d$, though exceptions may be found between closely related values. These could give an interesting insight in to the nature of noise on the device. Also, note that the minimum number of errors required to cause incorrect decoding is $\lceil d/2 \rceil$ (the number of flips on code qubits required to change (for odd $d$) or create ambiguity regarding (for even $d$) the majority. Since this number is the same for each odd $d$ and the even $d+1$ that follows it, there might not be a significant decrease in the logical error probability between these pairs.

For an additional test, note that the output of the code qubits alone is enough to perform decoding. This would not be true if many rounds of CNOTs and ancilla measurements were applied. In this case, the amount of noise built up on the code qubits would be too high for reliable decoding, and only with the history of ancilla measurement results can the original encoded value be deduced. However, in the case of the single ancilla measurement round as we consider, the noise level should be low enough to allow decoding using the code qubit results alone.

The result of this is that the code could satisfy the previously mentioned conditions (the code performing better than a single qubit, and ever better as $d$ increases) even if the CNOTs completely failed to occur. This would in no way be any proof-of-principle of quantum error correction, and so these conditions are clearly not sufficient to ensure success.

The use of the CNOTs and ancilla measurements in our experiment has both benefits and drawbacks. The former is due to the extra information that can be extracted about the errors that occur. The latter is due to the additional noise suffered by the code qubits due to imperfections in the CNOTs. For a proof-of-principle demonstration of quantum error correction, it must be shown that the benefits significantly outweigh the drawbacks.

To do this, we compare results for two types of decoding. One is `full decoding' in which the look-up tables of Eq. \ref{conditional} use the results from both the code and ancilla qubits (and therefore each $R$ is a string of $2d-1$ bits). The other is `partial decoding', in which only results from the code qubits are used to construct the look-up tables (and so each $R$ is a $d$ bit string). Since only the former benefits from the syndrome measurement round, since this records additional information about the errors on the ancilla qubits, it should lead to significantly lower logical error probabilities. By showing this, we would demonstrate that the effect of the CNOTs and ancilla measurements in extracting additional information truly has a powerful effect. Note that this form of test was originally proposed in \cite{winkler:17} for the surface code.

\section{Results}

Repetition codes of size between $d=3$ and $d=8$ were studied. For each, the lookup tables where populated using $8192$ samples. Note that this is significantly less than the total possible number of measurement outcomes, $2^{2d-1}$, for $d=8$ and not significantly greater than that for $d=6$ or $d=7$. Statistical inaccuracies in the lookup table can therefore be expected to affect the quality of the decoding. In order to get an idea of the extent of this effect, logical error probabilities for each case are calculated from $10$ different runs and a mean is taken. The standard deviation of these values is used as an estimate of error.

Simulated runs were also used to produce data that could be used for comparison. These runs were done in the same way as above, but only up to $d=6$.

The simulations also include artificially introduced noise, since they would be otherwise perfect. This was done using partial rotations about the X axis (which rotates between $\ket{0}$ and $\ket{1}$). The rotation angle depends on whether the qubit was initialized in state $\ket{0}$ or $\ket{1}$ at the encoding step. For the former, an angle of $\pi/20$ was used. For the latter, $\pi/10$ was used. This mimics the greater probability of $\ket{1}\rightarrow\ket{0}$ transitions in the real device. Noise was added to all qubits at three points: immediately after encoding, between the two rounds of CNOTs and immediately before measurement. This noise mechanism is chosen for its simple implementation, and is not expected to accurately reproduce the true noise processes in the real device. The values of $\pi/20$ and $\pi/10$ are chosen because they reproduce some values and features of the results from the real device.

Full details of the implementation for both the real device and the simulation, including source code and raw data, can be found at \cite{repetitionsource}.

\begin{figure}[t]
\begin{center}
{\includegraphics[width=\columnwidth]{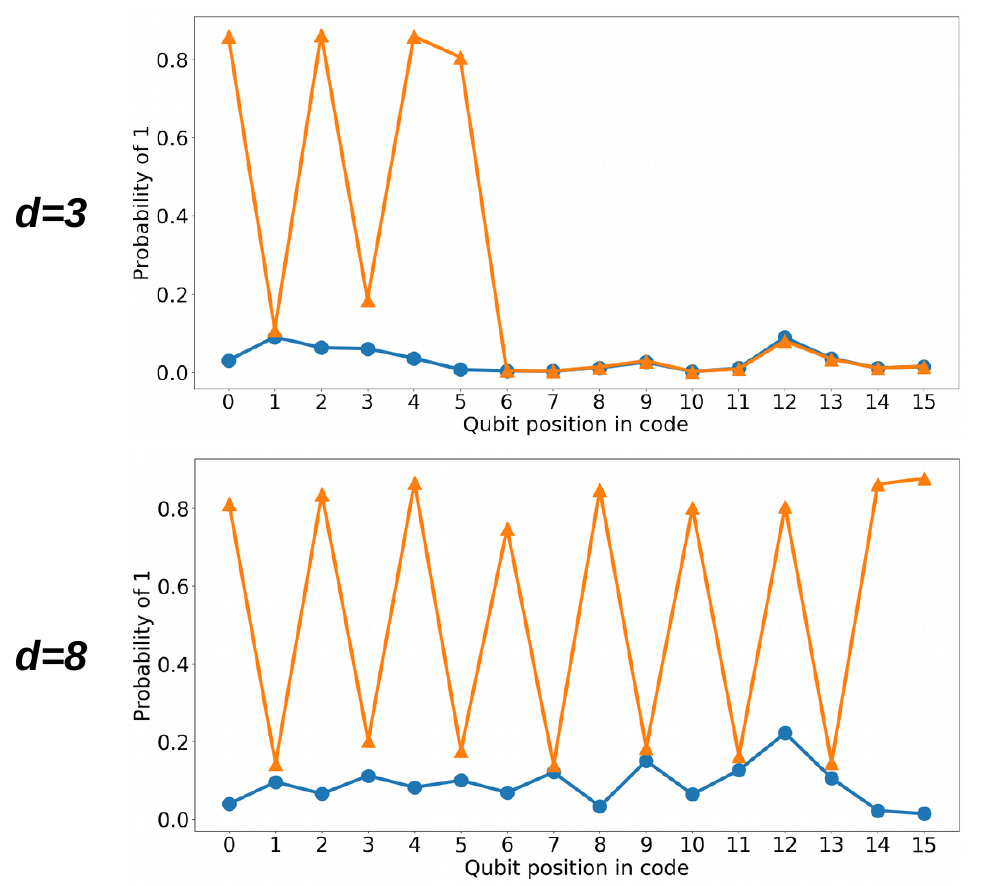}}
\caption{\label{singlequbits} The state of each qubit at the end of the process, characterized by the probability of the outcome $\ket{1}$ when measured. The extreme examples of $d=3$ and $d=8$ are shown. Results for the encoded bit value $0$ are shown in blue (circle markers), and those for $1$ are in orange (triangle markers).  Qubits alternate between code and ancilla qubits, finally ending in a code qubit at $2d-2$. Qubit $s$ is located at $2d-1$. Any remaining qubits are unused.
}
\end{center}
\end{figure}

\begin{figure}[t]
\begin{center}
{\includegraphics[width=\columnwidth]{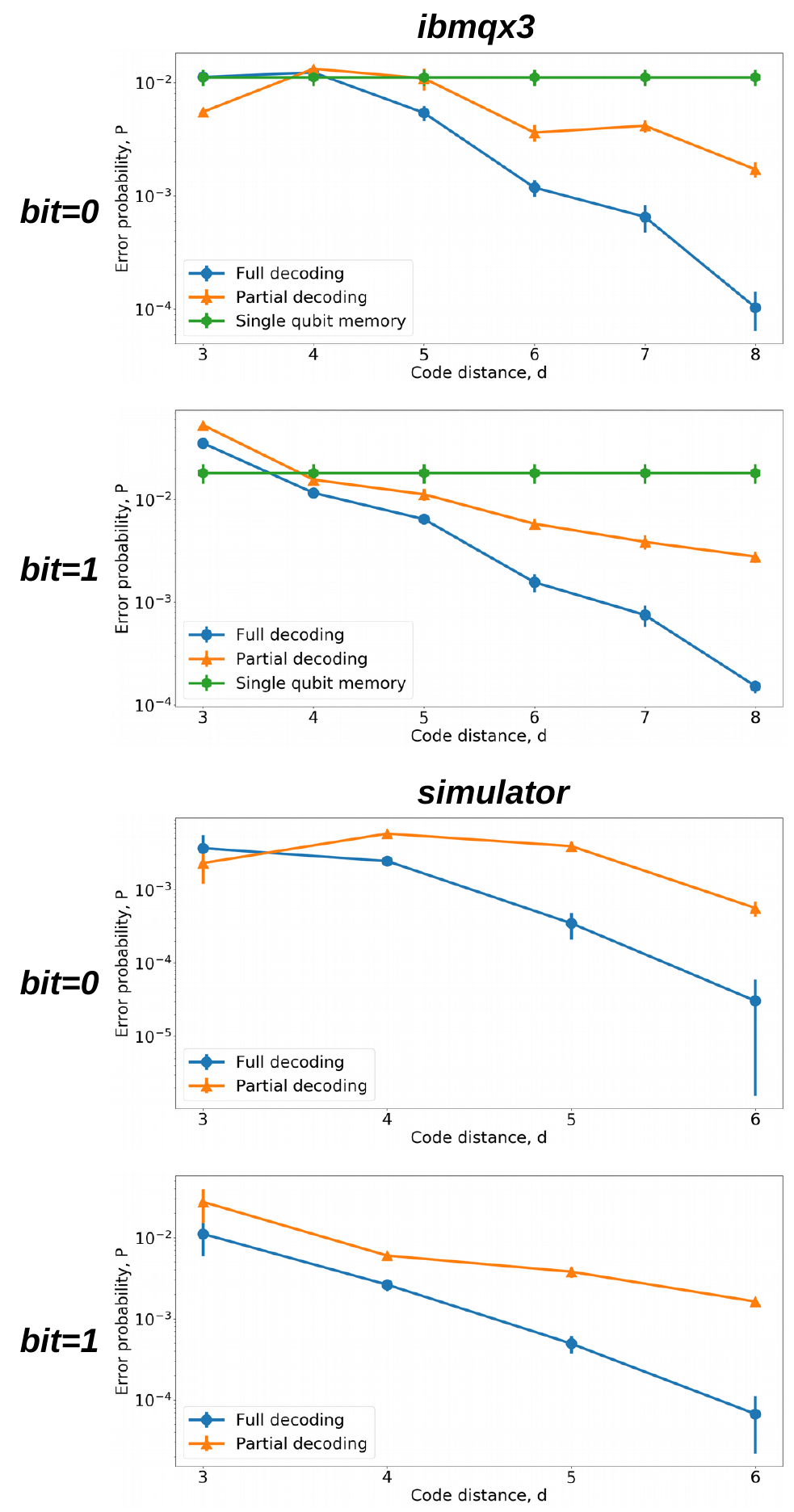}}
\caption{\label{logicalerrors} Probabilities of logical errors for both full and partial decoding. As a comparison, the minimum value of the single qubit memory from all code sizes is plotted across the graph.
}
\end{center}
\end{figure}

The results in Fig. \ref{singlequbits} show the state of each qubit at the end of the process. In the case of no noise, all qubit states should be $\ket{0}$, except for code qubits and qubit $s$ when the bit value $1$ encoded. The results show good agreement to these expectations. Qubits expected to be $\ket{0}$ typically have a fidelity of more than $90\%$, and those expected to be $\ket{1}$ have a fidelity of more than $80\%$. For the bit stored in the single qubit $s$, the lowest probability of a logical error is around $0.5\%$ for a stored $0$ and $10\%$ for a stored $1$. The values for the simulator show good agreement to those for the real device, though this is due only to the fact that the simulated noise parameters were chosen to obtain such an agreement.

The results in Fig. \ref{logicalerrors} show the logical error probabilities as a function of code distance. These probabilities are found to be much lower than the corresponding values for the qubit $s$ in the case of an encoded bit value of $0$. The effect is not so strong for the encoded $0$, however this is due to the fact that the noise tends cause qubits to decay towards $\ket{0}$ in all cases, which happens to be advantageous when $0$ is encoded.

The results show that the trend is indeed for the error probability to decrease with code distance, as we would expect. A notable exception is found at $d=6$ for an encoded bit value $0$. However, note that this is the first code that uses the qubit numbered $9$ in Fig. \ref{singlequbits}, which can be seen to be particularly noisy. Also, the error probability after full decoding is significantly less than that after partial decoding in almost all cases, which is again in line with expectations.

\subsection{Fit to an exponential decay}

To further analyse these trends, the data is fit to an exponential decay. The simplest possible model of the repetition code is used to avoid over fitting, or unstable fits. This model has a single free parameter, $p$, which corresponds to the probability of a bit flip error for each qubit. Decoding is done simply by using the majority voting of the code qubits The logical error probability for a given code distance $d$ is therefore the probability of such errors on at least half of the qubits within the code, and so decays exponentially with a factor of $\left[p/(1-p)\right]^{\ceil{d/2}}$. Note that the supremum here results in this factor being equal for each odd $d$ and the corresponding $d+1$, due to the lack of a clear majority in the even case.

\begin{figure}[t]
	\begin{center}
		{\includegraphics[width=\columnwidth]{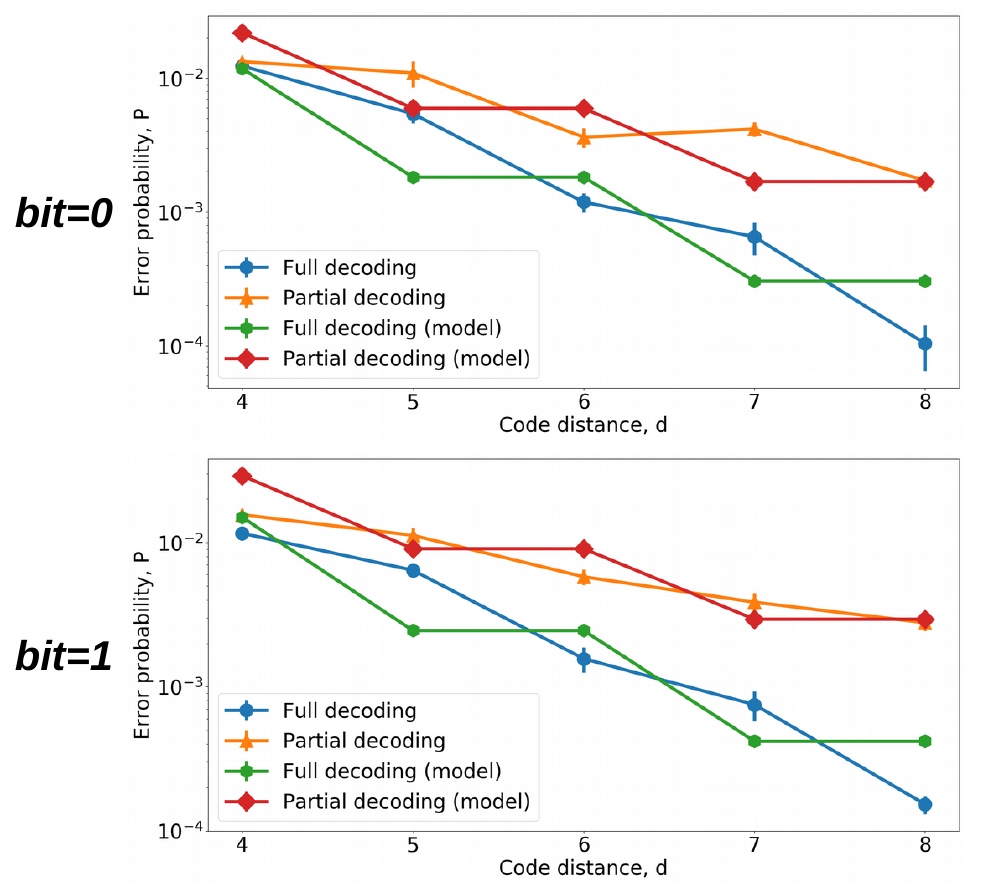}}
		\caption{\label{fit} Probabilities of logical errors for both full and partial decoding, as shown in Fig. \ref{logicalerrors}, with additional fit lines for an exponential decay. The results for $d=3$ have been omitted to reduce finite size effects.
		}
	\end{center}
\end{figure}

This simple model corresponds most closely to the case considered for partial decoding, and so is directly used to fit this data. The results are shown in Fig. \ref{fit}. The fitting was done using a least squares method on the logarithms of the logical error probabilities. The values for the fitting parameter were found to be $0.088\pm0.001$ for encoded $0$ and $0.102\pm0.001$ for encoded $1$. These values represent the combined effect of preparation, measurement and entangling gate errors, which are each measured to be on the order of $1\%$ using randomized benchmarking \cite{ibmqx3}. The fitting parameters are therefore certainly of the right order to represent their combined effect. However, the possibility that correlated noise can also form a significant contribution cannot be ruled out.

For the fitting of data from full decoding, let us consider two extremes. Firstly, consider the case for which all CNOTs and ancilla qubits are perfect. Full decoding would then effectively factor into two rounds of error correction, each of which can be modelled in the same way as partial decoding. We will refer to the values of the physical error probability for these rounds as $p_0$ and $p_1$, respectively. Clearly $p = p_0(1-p_1)+p_1(1-p_0)$, though we will use the approximation $p=p_0+p_1$ for simplicity.

The second extremal case is that for which the syndrome measurements result in no useful information being placed on the ancillas. This could either be due to the CNOTs being completely ineffective, or the ancillas being completely decohered. In this case, the ancilla results can be ignored. Full decoding could then be treated as a single round of partial decoding.

We will therefore fit the data for full decoding to two rounds of the simple model. This will be done by first assuming that one round has the error probability $p_0 = p$, found from the fit to the partial decoding data, and the other has no errors ($p_1=0)$. The fit will then optimize over all other possible $p_0$ and $p_1$ given the $p=p_0+p_1$ constraint.

The resulting $p_0$ and $p_1$ can then be used to determine the degree to which we can factorize the error correction into two rounds. If one of these probabilities is found to dominate, it would suggest that the effectiveness of the syndrome measurement approaches the worst case scenario described above.

A more even split would support the notion that the syndrome measurement round is indeed effective. However, it is not a complete proof. Similar results could occur if the CNOTs were ineffective, but suffered correlated noise. Full proof of effectiveness would therefore require a deeper understanding of the effects of noise in the system.

Fitting the data for the full decoding in this way yields $p_0=0.054\pm0.001$ and $p_1=0.034\pm0.001$ for stored $0$ and $p_0=p_1=0.051\pm0.001$ for stored $1$. These do show a clear sign of factoring into two distinct rounds, which strengthens the argument that the syndrome measurement round significantly increases the performance of the code.

The data can be seen to show a clear agreement with the exponential decay of the fit lines. However, the data and fitted lines do differ by a typical factor of around $2$. This is due to disagreement in the form of the even/odd effects. In some cases, such as partial decoding for an encoded $0$, the data and fit show opposite even/odd effects. This is likely due to biased noise changing the nature of the decoding, as will be discussed below. The even/odd effects in the data also appear to be less prominent, with partial decoding for an encoded $1$ as the clearest example. These differences could be accounted for by fitting to a more complex model that accounts for biased and correlated noise. However, this would increase the number of fitting parameters. A full study of these effects must therefore be deferred to future experiments with larger devices, such that more data points can be taken. Future experiments would also benefit from greater access to the raw data from devices, rather than the post-processed outputs of $0$ and $1$ as supplied currently.

\subsection{Analysis of finite size effects}

An exception to the decay of the logical error rate can be seen in Fig. \ref{logicalerrors} for $d=3$ and an encoded bit value of $0$. This occurs both for the real device and the simulator. It may seem counterintuitive that better results could come from ignoring some information. However, this effect is due to the way the probabilities are calculated  and the biased nature of the noise.

For example, consider an extreme case for which noise on the code qubits causes very strong relaxation, such that the code qubits always end in state $\ket{0}$ when $0$ was encoded, and almost always end in state $\ket{0}$ even when $1$ is encoded. Partial decoding using only these results would then lead to the decoding guessing that $0$ was encoded in almost all cases. There would then never be a logical error for an encoded $0$, but a logical error would be almost certain for an encoded $1$.

More informed decoding could be achieved using the ancilla results. If these show signs that extensive errors have occurred, it would be likely that the code qubits start in in state $\ket{1}$. Many cases for which a logical error would have occurred when using partial decoding could then be successfully decoded using this full decoding. However, for some cases, such as many measurement errors on ancilla qubits, it is possible for an encoded $0$ to be misidentified as a $1$. The use of a less biased decoder therefore may be more effective in overall, but it can be less effective for the specific case of an encoded $0$.

\begin{figure}[t]
\begin{center}
{\includegraphics[width=\columnwidth]{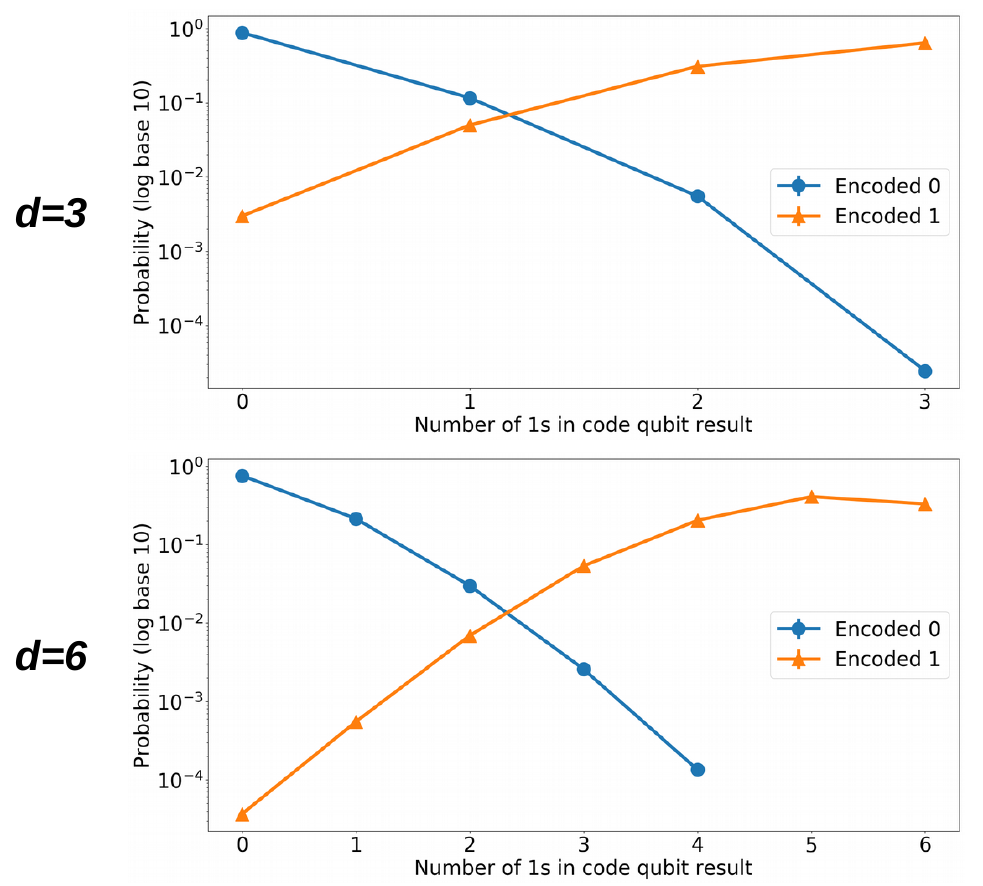}}
\caption{\label{decoder} Probabilities for different numbers of $1$s in the output of the code qubits.
}
\end{center}
\end{figure}

The reason for the effect in this specific case can be seen from Fig. \ref{decoder}. Here results are shown for $d=3$ and $d=6$. These graphs show how probable it is to get each possible number of the outcome $1$ in the output of the code qubits. For an encoded $0$ it is most probable to have a small number of $1$s, since each is an error deviating from the perfect output in which all are $0$. For an encoded $1$ the output is most likely to have a large number of $1$s, since each $0$ would be an error in this case.

For unbiased noise, the crossover between the curves for encoded $0$ and encoded $1$ would occur at $d/2$. Any output with less $1$s than this should therefore be decoded as a $0$, and any with more should be decoded as a $1$. For codes with even $d$, for which it is possible for the number of $1$s in the output to be the marginal value of $d/2$, the decoding can be chosen randomly in this case.

For the biased noise, as present in the real device and our simulations, the decoding can deviate from this simple majority voting. This can be seen in the results for $d=6$. The crossover point has shifted, resulting in $d/2$ no longer being the marginal case that would result for unbiased noise. Instead, having this number of $1$s in the output is recognized as being a strong indicator that the encoded bit value was $1$ and would be decoded accordingly. Similar effects occur for all other code distances, with the exception of $d=3$

For $d=3$ there is a similar shift of the crossover point. This shift is by approximately the same fraction of $d$ as in the $d=6$ case. However, the smaller nature of the code means that there is less freedom to alter the decoding accordingly. In fact, since finding two $1$s in the output is still slightly more likely to correspond to an encoded $0$, the optimal decoding will still correspond to majority voting. 

Nevertheless, outputs with two ones are found to be very close to being a marginal case, with encoded $0$ only being slightly more likely. Decoding these as $0$ in all cases will therefore lead to a unfair advantage for this encoded value, causing the feature found in Fig. \ref{logicalerrors}.

\section{Conclusions}

The logical error probabilities of the code sizes considered were found to agree with expectations in most cases, showing that current technology is certainly capable of achieving this simple example of quantum error correction on a relatively large scale. The exceptions found shed light on the nature of noise in the system, with the bias induced by relaxation being the dominant effect.

The quantum part of the code is the mapping of information about errors to ancilla qubits via controlled operations. This is a central technique of quantum error correction. Due to only a single round of ancilla measurements being used, it was possible to compare the effectiveness of decoding both with and without the ancilla results. This allowed direct insight into the effectiveness of the quantum part. It was found that it did indeed allow for significantly better results.

The next major goal of experimental quantum error correction is to build a logical qubit that can be stored as successfully as the logical bit here. The analysis used in this paper, such as the lookup table decoding and comparison of full and partial decoding, would be just as valid in that case \cite{winkler:17}. It would therefore be highly interesting to see corresponding results in future.

The most recent version of the Jupyter notebook containing source code for this project can be found at \cite{repetitionsource}. The raw data is also provided, allowing the analysis of this paper to be repeated and expanded upon. The version of the notebook on which this publication is based is included with the arXiv source files.

\section{Acknowledgements}

This work was supported by the Swiss National Science Foundation and the NCCR QSIT. JRW thanks Hanhee Paik for comments on the manuscript.

The IBM Quantum Experience was used to produce results for this work. The views expressed are those of the authors and do not reflect the official policy or position of IBM or the IBM Quantum Experience team.

\bibliography{refs}

\end{document}